\documentclass[twocolumn,prl,aps,twocolumn]{revtex4}
\usepackage[latin9]{inputenc}
\usepackage{amsmath}
\usepackage{amssymb}
\usepackage{graphicx}
\usepackage{makecell}

\usepackage[unicode=true,pdfusetitle,
 bookmarks=false,
 breaklinks=false,pdfborder={0 0 1},backref=false,colorlinks=false]
 {hyperref}
\hypersetup{
 bookmarksnumbered=false,bookmarksopen=false}

\makeatletter
\@ifundefined{textcolor}{}
{%
 \definecolor{BLACK}{gray}{0}
 \definecolor{WHITE}{gray}{1}
 \definecolor{RED}{rgb}{1,0,0}
 \definecolor{GREEN}{rgb}{0,1,0}
 \definecolor{BLUE}{rgb}{0,0,1}
 \definecolor{CYAN}{cmyk}{1,0,0,0}
 \definecolor{MAGENTA}{cmyk}{0,1,0,0}
 \definecolor{YELLOW}{cmyk}{0,0,1,0}
}

\usepackage{color}

\newcommand{\ehbar}{\hbar_{\mathrm{eff}}}

\@ifundefined{textcolor}{}{%
 \definecolor{BLACK}{gray}{0}
 \definecolor{WHITE}{gray}{1}
 \definecolor{RED}{rgb}{1,0,0}
 \definecolor{GREEN}{rgb}{0,1,0}
 \definecolor{BLUE}{rgb}{0,0,1}
 \definecolor{CYAN}{cmyk}{1,0,0,0}
 \definecolor{MAGENTA}{cmyk}{0,1,0,0}
 \definecolor{YELLOW}{cmyk}{0,0,1,0}
}

\usepackage{txfonts}

\makeatother

\begin{document}
\title{Quantization of Out-of-Time-Ordered Correlators in non-Hermitian Chaotic Systems}
\author{Wen-Lei Zhao}
\email{wlzhao@jxust.edu.cn}
\affiliation{School of Science, Jiangxi University of Science and Technology, Ganzhou 341000, China}

\begin{abstract}
This letter reports the findings of the late time behavior of the out-of-time-ordered correlators (OTOCs) via a quantum kicked rotor model with $\cal{PT}$-symmetric driving potential. An analytical expression of the OTOCs' quadratic growth with time is yielded as $C(t)=G
(K)t^2$. Interestingly, the growth rate $G$ features a quantized response to the increase of the kick
 strength $K$, which indicates the chaos-assisted quantization in the OTOCs' dynamics. The physics
behind this is the quantized absorption of energy from the non-Hermitian driving potential. This
discovery and the ensuing establishment of the quantization mechanism in the dynamics of quantum
chaos with non-Hermiticity will provide insights in chaotic dynamics, promising unprecedented observations in updated experiments.
\end{abstract}
\date{\today}

\maketitle

{\color{blue}\textit{Introduction.---}} More recently, the OTOCs have been proposed as effective indicators of various phenomena, for instance,  information scrambling~\cite{Maldacena16,Kuwahara21}, quantum butterfly effects~\cite{Maldacena16B,Shenker14,Lewis19}, many-body localization~\cite{Ray18,Rammensee18,Huang16}, quantum entanglement~\cite{Styliaris,Prakash20,Garttner18}, and dynamical phase transition~\cite{Dag19,Sahu19,Wangqian19,Heyl2018}. Therefore, it has attracted extensive investigations in diverse fields~\cite{Belyansky20,Mata2018,Lakshmi19,Rozenbaum17,Rozenbaum19,Rozenbaum20,Shenker15}. For holographic systems, it measures the spread of quantum information encoded in local degrees of freedom within the entire system~\cite{Jahnke19,Murthy19,Witten98}. Theoretical investigation demonstrates
the exponential growth of OTOCs with time, for which the rate, usually termed as quantum Lyapunov exponent, is  bounded by the inverse of the system's temperature~\cite{Maldacena16}. For non-holographic systems, it has been proved that the OTOCs are mathematically equivalent to the Loschmidt Echo, which provides a theoretical foundation for the application of the OTOCs as a probe of quantum chaos~\cite{Zurek20}. Indeed, the dynamics of the quantum OTOCs resembles its classical counterparts over the Ehrenfest time, both increasing exponentially with time and thus being the characteristics of the exponential instability of chaotic dynamics~\cite{Mata2018,Rozenbaum17,Carlos19}.
Such quantum-classical correspondence again strengthens the belief of exponential bound on the growth of OTOCs.
However, our recent work proves the
super-exponential growth of OTOCs as a result of the growth of chaoticity by the periodical modulation of nonlinear interaction in time domain~\cite{Zhao21}. More remarkably, the OTOCs even scale up linearly with the dimension of system, which implies the divergence of OTOCs for systems with infinitely large degrees of freedom, thus opening a new perspective on the dynamical behavior of chaotic systems.

The system's energy absorption from external potential  is a fundamental problem in quantum theory. Interestingly, nonlinearity, such as nonlinear Kerr effects and the mean-field approximation of interatomic  interaction, gives rise to the super-exponentially fast absorption of energy [$\propto \exp(e^{\gamma t})$] from non-Hermitian driving potential~\cite{Zhao20}.
During the past decades, the non-Hermitian physics had been under massive investigations in many fields of physics~\cite{Bender98,Moiseyev11}. As an important modification of quantum mechanics, non-Hermiticity has played versatile roles in describing the open quantum dynamics~\cite{Bender07}, the non-equilibrium relaxation problems~\cite{Verbaarschot}, and optical transport in lossy media~\cite{Xiao17,Freilikher94}. Novel phenomena induced by non-Hermiticity, for instance, topological insulator~\cite{Longwen21} and quantized momentum current~\cite{Zhao19}, have garnered research attentions both theoretically~\cite{Zhang19,Gong13,Ashida20} and experimentally~\cite{Ganainy18,Zhang16,Li19}. Previously, we even found that the chaotic dynamics of complex trajectories in non-Hermitian systems can be characterized by the exponential behavior of OTOCs~\cite{Zhao20B}.
In this context, the quantum dynamics of non-Hermitian chaotic systems demands urgent investigation.

In this letter, we investigate the late time dynamics of OTOCs via a kicked rotor system with $\cal{PT}$-symmetric kicking potential. We find that the OTOCs exhibit the quadratic growth versus time, i.e., $C(t)=G(K)t^2$. Remarkably, the growth rate $G(K)$ displays the quantized response to the $\cal{PT}$-symmetric drive, namely, it grows in a quantized way with the increase of the kicking strength $K$. We theoretically prove the equivalence between the OTOCs and the mean square of momentum, i.e., $C(t)=\nu\langle p^2(t)\rangle$. It is then clear that the quantization of OTOCs' growth rate is rooted in the quantized absorption of energy from the external non-Hermitian potential.
Our finding of the relation between OTOCs and energy diffusion is of particular significance in the interdisciplinary  fields ranging from many-body quantum chaos to quantum holography.
The novel quantization in OTOCs' dynamics may be within reach of up-to-date experiments, since non-Hermitian systems have been recently realized with atom-optics setting~\cite{Kreibich16,Kreibich13,Kreibich14} and photonics in waveguides~\cite{Xia21,Mukherjee20}. The quantized response of OTOCs to external driving potential  provides a theoretical foundation of engineering experimentally the information scrambling and the quantum states transportation.

{\color{blue}\textit{Model and results.---}}
The Hamiltonian of the $\cal{PT}$-symmetric kicked rotor ($\cal{PT}$KR) reads
\begin{equation}\label{Hamil}
{\rm{H}}=\frac{p^2}{2} + K\left[\cos(\theta)+ i\lambda \sin(\theta)\right]\sum_n \delta (t-t_n)\;,
\end{equation}
where the angular momentum operator $p=-i\ehbar\partial/\partial \theta $ and $\theta$ is the angle coordinate. The parameters $\ehbar$, $K$ and $\lambda$ are the effective Planck constant, the strength of the real parts and the imaginary parts of the kicking potential, respectively~\cite{Zhao19,Zhao20B,Longhi17}. The $t_n$($=1,2\ldots$) indicates the number of kicks. All variables are properly scaled and thus in dimensionless units.
The evolution of a quantum state from $t=t_n$ to $t_{n+1}$ is given by $|\psi(t_{n+1})\rangle = U |\psi(t_{n})\rangle$,
where the Floquet operator takes the form $U=\exp({-i p^2}/{2\ehbar})\exp\left\{-i{K}[\cos(\theta)+i\lambda \sin(
\theta)]/{\ehbar}\right\}$.
On the basis of angular momentum operator $p|\phi_n\rangle =p_n |\phi_n\rangle$ with $p_n=n\ehbar$, an arbitrary state can be expanded as $|\psi\rangle = \sum_{n} \psi_n |\phi_n\rangle$.

The OTOCs are defined by $C(t_n)  = - \langle [A(t_n), B(t_0)]^{2}\rangle$, where the $A(t_n)=U^{\dagger}(t_0,t_n)AU(t_0,t_n)$ and $B(t_0)$ are operators in  Heisenberg picture, and $\langle \cdot \rangle$ indicates the average over an initial state. Without loss of generality, we consider the case with $A=p$ and $B=\theta$, namely
\begin{align}\label{OTOCPX0}
C(t_n) & = - \left\langle [p(t_n), \theta(t_0)]^{2} \right\rangle \;.
\end{align}
Our main finding is that the OTOCs are the quadratic function of time
\begin{align}\label{OTOCPX}
C(t_n) = G t_n^2 \;,
\end{align}
where the growth rate displays the quantized growth with an increasing $K$, i.e.,
\begin{equation}\label{QunGrowth}
G (K) =  \nu (2m\pi)^2\quad \text{for}\quad K\in (2m\pi-\pi, 2m\pi+\pi)\;,
\end{equation}
with $\nu$ being a constant and $m\geq 1$ the integer.
In numerical investigations, the initial state is taken as a Gaussian wavepacket $\psi(t_0,\theta) = ({\sigma}/{\pi})^{1/4} \exp(-\sigma\theta^2/2)$ with $\sigma=10$. The reason lies in that our model is a kind of non-holographic dual system featuring infinite energy capacity from unbounded heating, absent of the well-defined thermal states. In contrast, many-body systems in previous investigations were mostly holographic dual ones, whose average of OTOCs in Eq.~\eqref{OTOCPX0} could be readily taken for the thermal states.

\begin{figure}[t]
\begin{center}
\includegraphics[width=8.0cm]{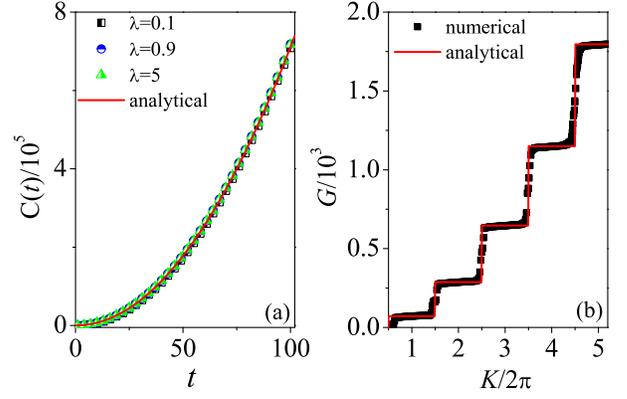}
\caption{(a) Time dependence of $C(t)$ with $\lambda=0.1$ (squares), $0.9$ (circles) and $5$ (triangles). Red line indicates our theoretical prediction in Eq.~\eqref{OTOCPX}. The parameters are $K=2\pi$ and $\ehbar=0.1$. (b) The growth rate $G$ of $C(t)$ versus $K$ with $\lambda=0.9$ and $\ehbar=0.1$. Red line denotes theoretical prediction in Eq.~\eqref{QunGrowth} with $\nu\approx 1.8$. \label{QunOTOC}}
\end{center}
\end{figure}

We concentrate on the chaotic dynamics in the broken $\cal{PT}$-symmetry phase ensured by the condition  $K\lambda/\ehbar\gg 1$~\cite{Zhao19,Longhi17}.
Figure~\ref{QunOTOC}(a) shows the evolution of $C(t_n)$, where one can find the perfect consistence between numerical results and the analytic prediction in Eq.~\eqref{OTOCPX}. Closer observations show that the time dependence of $C(t_n)$ for different $\lambda$ has almost no difference, which means that the growth rate $G$ does not change with $\lambda$.  As a further step, we numerically investigate the growth rate $G$ for different $K$. Interestingly, the value of $G$ demonstrates the growth in a quantized way with the increase of $K$ [see Fig.~\ref{QunOTOC}(b)], which is in good agreement with our theoretical prediction in Eq.~\eqref{QunGrowth}. Our investigation of the quantized response of OTOCs to $\cal{PT}$-symmetric driving potential is an important extension of Floquet engineering in non-Hermitian chaotic systems.

{\color{blue}\textit{Theoretical analysis.---}}
The OTOCs can be decomposed as
\begin{align}\label{Decomposition}
C(t_n) & =  C_1(t_n) + C_2(t_n) - 2{\rm{Re}}[C_3(t_n)]\;,
\end{align}
where we define the first term in the right as
\begin{align}\label{FPart}
C_{1}(t_n)& \mathrel{\mathop:}= \langle p^{\dagger}(t_n)\theta^2p(t_n) \rangle=\langle \psi_R(t_0) | \theta^2 |\psi_R(t_0)\rangle\;,
\end{align}
the second term as
\begin{align}\label{SPart}
C_{2}(t_n) &\mathrel{\mathop:}= \langle \theta p^{\dagger}(t_n)p(t)\theta \rangle =\langle \varphi_R(t_0)|\varphi_R(t_0)\rangle\;,
\end{align}
and the third one as
\begin{align}\label{TPart}
C_{3}(t_n) &\mathrel{\mathop:}= \langle p^{\dagger}(t_n)\theta p(t_n)\theta \rangle=\langle \psi_R(t_0)|\theta |\varphi_R(t_0)\rangle\;,
\end{align}
with $|\psi_R(t_0)\rangle = U^{\dagger}(t_0,t_n)pU(t_0,t_n)|\psi(t_0)\rangle$ and $|\varphi_R(t_0)\rangle =U^{\dagger}(t_0,t_n)pU(t_0,t_n)\theta|\psi(t_0)\rangle$ being the states at the end of time reversal.
Both $C_1$ and $C_2$ are termed as two-point correlators, and $C_3$ is the four-point correlator of OTOCs.

\begin{figure}[t]
\begin{center}
\includegraphics[width=8.5cm]{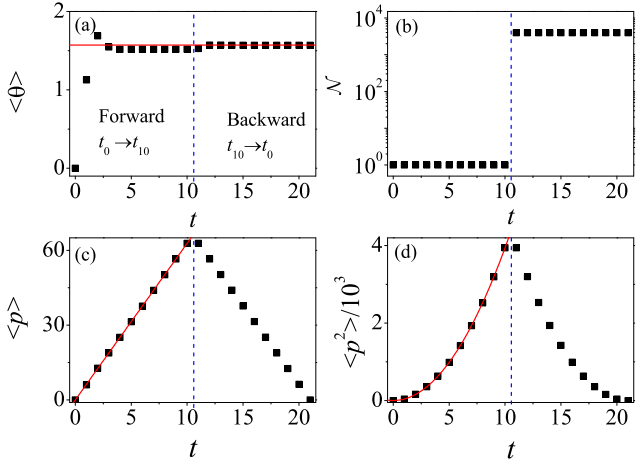}
\caption{Time evolution of  $\langle \theta\rangle$ (a), $\cal{N}$ (b), $\langle p\rangle$ (c), and $\langle p^2\rangle$ (d). In (a): Red solid line marks $\theta_c = \pi/2$. From (b), one can see that, for $t>t_{10}$, the norm of quantum states equals to the mean square of momentum at the time $t_{10}$, i.e., $\mathcal{N}(t) = \langle \psi(t_{10})| p^2|\psi(t_{10})\rangle $. For $t<t_{10}$, the norm is unity $\mathcal{N}(t) =1$. In (c): Red solid line denotes analytical prediction $\langle p\rangle = D t$ in Eq.~\eqref{LinAccelera}. In
(d): Red solid line indicates theoretical prediction $\langle p^2\rangle = D^2 t^2$ in Eq.~\eqref{BlsDiffu}.
The results in (c) and (d) display the perfect time reversal. Blue dashed lines are auxiliary lines. The parameters are $K=2\pi$, $\lambda=0.9$, and $\ehbar=0.1$.  \label{TMEReversal}}
\end{center}
\end{figure}

\begin{figure}[t]
\begin{center}
\includegraphics[width=8.5cm]{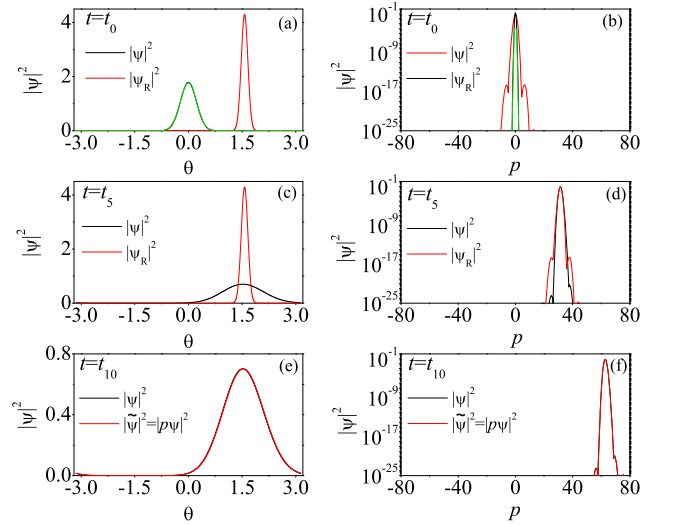}
\caption{Distributions in real (left panels) and momentum (right panels) space. In (a)-(d): black and red lines
indicate the distribution of the states at the forward $\psi(t)$  and backward $\psi_R(t)$ time evolution, with $t=t_0$ (top panels) and $t_5$ (middle panels). Green lines denote the initial Gaussian wavepacket. Bottom panels: Distribution of the states in real (e) and momentum (f) space with $t=t_{10}$. Black and red lines indicate the state $\psi(t_{10})$ and $\tilde{\psi}(t_{10})=p\psi(t_{10})$, respectively. For comparison, all of the states are normalized to unity. Other parameters are the same as in Fig.~\ref{TMEReversal}.
\label{distributions}}
\end{center}
\end{figure}

Equation~\eqref{FPart} demonstrates that at a specific time $t_n$, the term $C_1(t_n)$ is just the mean square of angle coordinate taken for the state $|\psi_R(t_0)\rangle$ at the end of time reversal.
We numerically investigate both the mean value $\langle \theta(t_j)\rangle = \langle\psi (t_j)|\theta|\psi (t_j)\rangle/{\mathcal{N}(t_j)}$
and the norm $\mathcal{N}(t_j) = \langle\psi (t_j)|\psi (t_j)\rangle$ during the forward ($t_0 \rightarrow t_n$) and backward ($t_n \rightarrow t_0$) time evolution.
Our numerical results show that the value of $\langle \theta\rangle$ increases rapidly from zero to a fixed value $\theta_c =\pi/2$ during the forward time evolution [e.g., $t_0\rightarrow t_{10}$ in Fig.~\ref{TMEReversal}(a)]. During the backward evolution, it exhibits the perfect recovery, i.e., $\langle \theta\rangle = \theta_c=\pi/2$, except for the last two kicks. At the end of the time reversal, the value of $\langle \theta \rangle$  equals to $\pi/2$ [see the $\psi_R(t_0)$ in Fig.~\ref{distributions}(a)], while that of the initial time $t=t_0$ is zero [see the $\psi(t_0)$ in Fig.~\ref{distributions}(a)]. The difference results from the effect of the $\cal{PT}$-symmetric potential. Remember that the maximum value of the Floquet operator of the imaginary kicking potential, $U_K^{\lambda}=\exp[K\lambda\sin(\theta)/\ehbar]$ corresponds to $\theta_c=\pi/2$, then the action of $U_K^{\lambda}$ on a quantum state will greatly increase the amplitude of the quantum state at $\theta_c=\pi/2$ provided $K\lambda/\ehbar \gg 1$.
In condition that the wavepacket $|\psi_R(t_0)\rangle$ is extremely localized at $\theta_c$, it is straightforward to get the mean value $\langle\psi_R(t_0)| \theta^2|\psi_R(t_0) \rangle \approx \alpha \mathcal{N}_{\psi_R}(t_0)$, where the factor $\alpha$ equals to $\theta_c^2$, i.e., $\alpha\approx \theta_c^2$
and $\mathcal{N}_{\psi_R}(t_0)$ is the norm of the state $|\psi_R(t_0) \rangle$.

The backward evolution starts from the state $|\tilde{\psi}(t_n)\rangle=p|\psi(t_n)\rangle$, which is generated by the operation of $p$ on the state $|\psi(t_n)\rangle$~\cite{Supple}. During the time reversal from $t_n$ to $t_0$, we take the normalization for the time-evolved state so that the norm $\tilde{\mathcal{N}}_{\psi}(t_n)=\langle \tilde{\psi}(t_n)| \tilde{\psi}(t_n)\rangle$ remains as a constant, i.e., $\mathcal{N}_{\psi_R}(t_0)= \tilde{\mathcal{N}}_{\psi}(t_n)$ [see Fig.~\ref{TMEReversal}(b)]~\cite{Supple}.
Taking $\tilde{\mathcal{N}}_{\psi}(t_n)=\langle \psi(t_n)|p^2| \psi(t_n)\rangle$ into account, we get the relation $C_1(t_n)= \alpha \langle \psi(t_n)|p^2| \psi(t_n)\rangle$.
Interestingly, during the forward evolution from $t_0$ to $t_n$, the expectation value of $\langle p\rangle$ increases linearly as [e.g., $t_0\rightarrow t_{10}$ in Fig.~\ref{TMEReversal}(c)]
\begin{equation}\label{LinAccelera}
\langle p\rangle =D t_n\;,
\end{equation}
where the growth rate shows the quantization with respect to $K$,
\begin{equation}\label{QUNACCEL}
D=2m\pi\quad \text{for} \quad K\in (2m\pi-\pi,2m\pi+\pi)\;,
\end{equation}
with $m\geq 1$ being an integer~\cite{Zhao19}. Such unidirectional transport of wavepackets can also be seen from the momentum distribution in Figs.~\ref{distributions}(b),~\ref{distributions}(d) and~\ref{distributions}(f), where one can find
a soliton with a center $p_c$ moving to the positive direction as time evolves, i.e., $p_c\approx D t$.

\begin{figure}[t]
\begin{center}
\includegraphics[width=8.5cm]{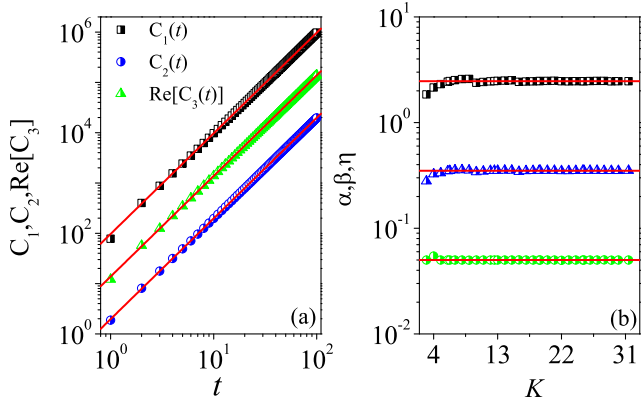}
\caption{(a) Time evolution of $C_1(t)$ (squares), $C_2(t)$ (circles) and $\textrm{Re}[C_3(t)]$ (triangles) with $K=2\pi$. Red lines (symbols) indicate the theoretical prediction (numerical results) in Eqs.~\eqref{MSQMTR2}[$C_1(t)$],~\eqref{SEDOTOC2}[$C_2(t)$] and~\eqref{TPart4}[$C_3(t)$], respectively. (b) The value of $\alpha$ (squares), $\beta$ (circles), and $\eta$ (triangles) versus $K$. From top to bottom, red lines indicate $\alpha \approx \pi^2/4$, $\eta\approx 0.35$, and $\beta \approx 1/2\sigma$, respectively. The parameters are $\lambda=0.9$ and $\ehbar=0.1$.\label{ThreeTerm}}
\end{center}
\end{figure}

In condition that the wavepacket is sufficiently narrow in momentum space, we get the approximation of the mean energy
\begin{equation}\label{BlsDiffu}
\langle p^2\rangle \approx p_c^2\approx D^2 t_n^2\;,
\end{equation}
which is the typical phenomenon of ballistic diffusion [see Fig.~\ref{TMEReversal}(d)]. Accordingly, the
time dependence of $C_1(t_n)$ obeys the law
\begin{equation}\label{MSQMTR2}
C_1(t_n)= \alpha D^2 t_n^2\;.
\end{equation}
Our theoretical analysis is verified by numerical results as shown in Fig.~\ref{ThreeTerm}(a). Moreover, we numerically investigate the value of $\alpha$ for different $K$. Interestingly, it is a constant, i.e., $\alpha\approx \theta_c^2$ as $K$ varies [see Fig.~\ref{ThreeTerm}(b)].

We proceed to derive the time-dependence of $C_2(t_n)$. Equation~\eqref{SPart} means that the $C_2(t_n)$ is just the norm of the state $|\varphi_R(t_0)\rangle$, i.e., $C_2(t_n)=\mathcal{N}_{\varphi_R}(t_0)=\langle \varphi_R(t_0)|\varphi_R(t_0)\rangle$.
It is worth noting that during the time reversal ($t_n\rightarrow t_0$), we have taken the normalization after each kick. So there is the equivalence $\mathcal{N}_{\varphi_R}(t_0)=\tilde{\mathcal{N}}_{\varphi}(t_n)$,
where $\tilde{\mathcal{N}}_{\varphi}(t_n)$ is the norm of the state $|\tilde{\varphi}(t_n)\rangle$($=p|\varphi(t_n)\rangle$)~\cite{Supple}. It is apparent that this norm is just the mean square of momentum at the time $t=t_n$, i.e., $\tilde{\mathcal{N}}_{\varphi}(t_n)=\langle \varphi(t_n)|p^2|\varphi(t_n)\rangle$.
According to the above analysis, the quantum mean energy diffuses ballistically, i.e., $\langle p^2(t_n)\rangle= \mathcal{N}_{\varphi_0} D^2t_n^2$ [see Eq.~\eqref{BlsDiffu}],
where $\mathcal{N}_{\varphi_0}$ is the norm of the initial state $|\varphi(t_0)\rangle$($=\theta|\psi(t_0)\rangle$)~\cite{Supple}. For an initial Gaussian wavepacket $\psi(t_0,\theta)=({\sigma}/{\pi})^{1/4} \exp(-{\sigma \theta^2}/{2})$, it is straightforward to get $\mathcal{N}_{\varphi_0}= \langle \varphi(t_0)|\varphi(t_0)\rangle=\langle \psi(t_0)|\theta^2|\psi(t_0)\rangle={1}/{2\sigma}$.
Then, the time-dependence of $C_2(t_n)$ has the expression
\begin{equation}\label{SEDOTOC2}
C_2(t_n)= \beta D^2 t_n^2 \qquad \text{with} \qquad \beta= \mathcal{N}_{\varphi_0}=\frac{1}{2\sigma}\;,
\end{equation}
which is in good agreement with our numerical results [see Fig.~\ref{ThreeTerm}(a)]. The factor $\beta$ has verified the relation $\beta =1/2\sigma$, no matter what the $K$ is [see Fig.~\ref{ThreeTerm}(b)].

\begin{figure}[t]
\begin{center}
\includegraphics[width=7.5cm]{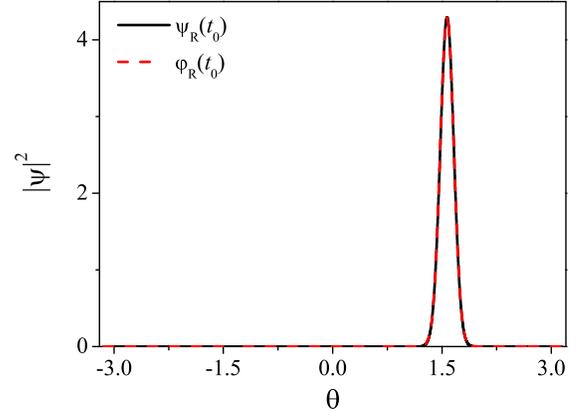}
\caption{Comparison of density distribution between $\psi_R(t_0)$ (black solid curve) and $\varphi_R(t_0)$ (red-dashed curve) at the end of time reversal. For comparison, both the $\psi_R(t_0)$ and $\varphi_R(t_0)$ have been normalized to unity, i.e., $\langle \psi_R(t_0)|\psi_R(t_0)\rangle = \langle \varphi_R(t_0) | \varphi_R(t_0)\rangle=1$.
Other parameters are the same as in Fig.~\ref{TMEReversal}. \label{DisTREINI}}
\end{center}
\end{figure}

As a further step, we analytically estimate the four-points correlator $C_3(t_n)=\langle \psi_R(t_0)|\theta |\varphi_R(t_0)\rangle$ [see Eq.~\eqref{TPart}], which in general is complex since the two states $|\psi_R(t_0)\rangle$ and $|\varphi_R(t_0)\rangle$ are different. However, with the only exception being the norm, the two states in our system are almost twin-like solitons centered at $\theta_c=\pi/2$ (see Fig.~\ref{DisTREINI}). Therefore, the term $C_{3}$ can be approximately regarded as the mean angle taken with respect to $|\psi_R(t_0)\rangle$ (or $|\varphi_R(t_0)\rangle$).  A rough estimation immediately yields $C_{3}(t_n) \approx \eta \sqrt{\mathcal{N}_{\psi_R}(t_0)}\sqrt{\mathcal{N}_{\varphi_R}(t_0)}$, where the factor $\eta$ is related to the center $\theta_c=\pi/2$ of the wavefunction $|\psi_R(t_0)\rangle$ (or $|\varphi_R(t_0)\rangle$), with $\mathcal{N}_{\psi_R}(t_0)$ and $\mathcal{N}_{\varphi_R}(t_0$) indicating the norm of the $|\psi_R(t_0)\rangle$ and $|\varphi_R(t_0)\rangle$, respectively.
According to the above analysis, the norm of $|\psi_R(t_0)\rangle$ equals to the mean square of momentum at the time $t=t_n$, i.e., $\mathcal{N}_{\psi_R}(t_0)=\langle \psi(t_n)|p^2|\psi(t_n)\rangle=D^2 t_n^2$ [see Fig.~\ref{TMEReversal}(b)].
For the state $|\varphi_R(t_0)\rangle$, its norm is $\mathcal{N}_{\varphi_R}(t_0)=\langle \varphi(t_n)|p^2|\varphi(t_n)\rangle=\mathcal{N}_{\varphi_0}D^2 t_n^2$ with $\mathcal{N}_{\varphi_0}=1/2\sigma $. Then, it is straightforward to get the time-dependence four-point correlator
\begin{align}\label{TPart4}
C_{3}(t_n) &= \eta D^2 t_n^2\;,
\end{align}
where the factor $\mathcal{N}_{\varphi_0}=1/2\sigma$ has been included in the coefficient $\eta$.
We numerically investigate the $C_{3}(t_n)$ and find that its imaginary part $\textrm{Im}[C_{3}]$ is much smaller than its  real part $\textrm{Re}[C_{3}]$, which implies that the $C_3$ is roughly real. Moreover, the time-dependence of $\textrm{Re}[C_{3}]$ agrees well with our theoretical prediction in Eq.~\eqref{TPart4} [see Fig.~\ref{ThreeTerm}(a)]. Our numerical results of the dependence of $\eta$ on $K$ demonstrate that it remains almost the same as $K$ varies, hence a constant [see Fig.~\ref{ThreeTerm}(b)].

Combining Eqs.~\eqref{Decomposition},~\eqref{MSQMTR2},~\eqref{SEDOTOC2}, and~\eqref{TPart4} yields the law of the time-dependence of OTOCs
\begin{equation}\label{MSQMTR3}
C(t_n)= G t_n^2\quad \text{with}\quad  G=\nu D^2 \;,
\end{equation}
where the prefactor $\nu=\alpha + \beta -2\eta$.
It is clear that the quantized increase of $G$ with $K$ is due to the quantized growth of $D$ with $K$ [see Eq.~\eqref{QUNACCEL}], which has been reported in our previous work~\cite{Zhao19}. Our theoretical prediction of the quantization of OTOCs is in good agreement with numerical results (see Fig.~\ref{QunOTOC}). The quadratic time-dependence of OTOCs is rooted in the ballistic diffusion $\langle p^2(t)\rangle = D^2 t^2$ in the momentum space, due to the mathematical equivalence $C(t)=\nu \langle p^2(t)\rangle$. The relation between quantum scrambling and chaotic diffusion is an elusive issue in the fields of black hole physics and quantum chaos~\cite{Lewis19}. Our theoretical finding provides a strong evidence that, in a kind of Floquet systems with $\cal{PT}$-symmetric external potential, the OTOCs of a generic form can be used to identify the energy-heating. It is particularly significant for the process of thermalization of non-Hermitian chaotic systems.

\begin{figure}[t]
\begin{center}
\includegraphics[width=8.5cm]{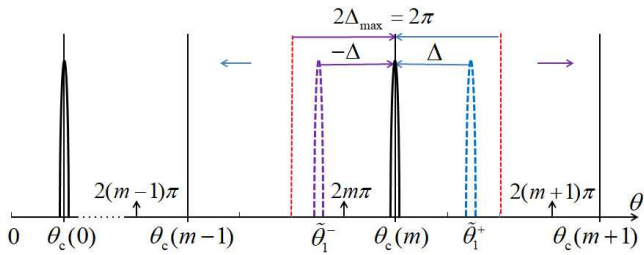}
\caption{Schematic diagram for the movement of a soliton. Solid lines mark the angle coordinate
 $\theta_c(m)=2m\pi+\pi/2$ with $m=0,1\ldots$. Red dashed lines denote the middle points between $\theta_c(m-1)$ and $\theta_c(m)$. Up arrows indicate $\theta=2m\pi$.
 Horizon arrows mark the direction of the motion of a soliton. The soliton is initially concentrated at $\theta_0=\theta_c(0)=\pi/2$ with $p_0=0$. The kick strength is $K=2m\pi \pm \Delta$. After a kick, the real part of kick potential drives the soliton to the position  $\tilde{\theta}_1^+=\theta_c(m)+\Delta$ for $K=2m\pi + \Delta$, and to $\tilde{\theta}_1^-=\theta_c(m)-\Delta$ for $K=2m\pi - \Delta$. If the soliton is in the region $(\theta_c(m)-\Delta_{max},\theta_c(m)+\Delta_{max})$ with $\Delta_{max}=\pi$, the effect of imaginary part of kick potential causes the movement of the soliton to $\theta_c(m)$. Therefore, the momentum is actually $p=\theta_c(m)-\theta_0=2m\pi$. The acceleration is $D=p-p_0=2m\pi$ with $K\in (2m\pi-\pi,2m\pi+\pi)$.\label{claPicture}}
\end{center}
\end{figure}

For completeness, we briefly explain the quantized acceleration of the momentum current $\langle p\rangle$. In our system, the real part of $\cal{PT}$-symmetry-kick provides the driving force $F=K\sin(\theta)$, while its imaginary part has a kind of dissipation effects to produce soliton with a center $\theta_c=\pi/2$($+2m\pi$) [see Fig.~\ref{distributions}(a), Fig.~\ref{distributions}(c), and Fig.~\ref{distributions}(e)]. Without loss of generality, we consider a soliton initially centered at $(\theta_0=\theta_c,p_0=0)$, and assume that $K=2m\pi \pm \Delta$. At a kick, it experiences a kicking force of strength $F (\theta_c) = K\sin(\theta_c)=K$, so its momentum changes to be $\tilde{p}_1=p_0+F=2m\pi \pm \Delta$. Consequently, the angle coordinate is $\tilde{\theta}_1=\theta_0 + \tilde{p}_1=\theta_c+2m\pi \pm \Delta$.
At first glance, the acceleration of a soliton is
$D=\tilde{p}_1-p_0 =K$, which marks the continuous growth of $D$ without the quantization with a varying $K$. In fact, the fantastic phenomenon of quantized acceleration is due to the modification of the dissipation effects of the imaginary part of the $\cal{PT}$-symmetry-kick on the classical acceleration.
Note that the dissipation effects always give rise to a soliton at $\theta_c=\pi/2$($+2m\pi$), thus the deviation $\Delta$ of $\tilde{\theta}_1$ from $\theta_c+2m\pi$ should be reduced, i.e., the coordinate is $\theta_1=\tilde{\theta}_1 \mp\Delta=\theta_c+2m\pi$ (see Fig.~\ref{claPicture}). This means that the momentum $p_1=\theta_1-\theta_0=2m\pi$($= \tilde{p}_1\mp \Delta$), thus the acceleration $D=p_1-p_0=2m\pi$.
According to the above analysis, the dissipation effects reduce the difference ($\Delta$) of the angle coordinate from $\theta_c+2m\pi$. It is then reasonable to believe that the dissipation effects change the position of a soliton to be $\theta_c+2m\pi$ for $\Delta < \pi$, and $\theta_c+2(m+1)\pi$ for $\Delta > \pi$.
A natural estimate of $\Delta$ is a half of the distance from $2m\pi + \theta_c$ to $2(m+1)\pi + \theta_c$, namely, $\Delta = \pi$.
Then, we get the quantized growth of $D$, i.e., $D=2m\pi$ with $K\in (2m\pi-\pi, 2m\pi+\pi)$, which has been numerically verified in Ref.~\cite{Zhao19}.

Finally, we give an explanation for the time reversal as shown in Fig.~\ref{TMEReversal}. The backward evolution starts from the state $|\tilde{\psi}(t_n)\rangle$, which is generated by the operation of $p$ on the state $|\psi(t_n)\rangle$ at the time $t=t_n$~\cite{Supple}. Our investigation shows that the density distributions for $|\tilde{\psi}(t_n)\rangle$ and $|\psi(t_n)\rangle$ have very slight differences [see Figs.~\ref{distributions}(e) and (f) for $t=t_{10}$], which means that the action of $p$ does almost  not change the coordinate of the quantum particle in the phase space. So, at the beginning of time reversal, the quantum particle just changes the direction of momentum.
During the time reversal process, the $\cal{PT}$-symmetric potential yields a soliton
with the center $\theta_c=\pi/2$ which is the same as that of the soliton at the forward time evolution [see Fig.~\ref{distributions}(c) for $t=t_5$]. Therefore, a force will be exerted on the particle at each time reversal kick, which, compared with the forward evolution, features the same strength but reversed direction.
 As a consequence, the quantum particle will recover the original trajectory ($\langle \theta\rangle,\langle p\rangle$) step by step [see the expectation value in Figs.~\ref{TMEReversal}(a) and (c)].

{\color{blue}\textit{Conclusion and prospects---}}
In this letter, we investigate the dynamics of OTOCs in a $\cal{PT}$KR model numerically and  analytically, with research emphasis on the
later-time behavior of $C(t)$ in the strong chaotic situation $K\gg 1$. Quadratic growth with time, i.e., $C(t)=Gt^2$ is found on the condition that $K\lambda/\ehbar \gg 1$. Quantized response of OTOCs to the external driving potential is characterized by the staircase increment of $G$ with the increase of $K$, i.e., $G=\nu (2m\pi)^2$ with $K\in (2m\pi-\pi,2m\pi+\pi)$ ($m=1,2\ldots$), which is a clear evidence of chaos-assisted quantization.
We theoretically establish the relation between OTOCs and energy diffusion, i.e., $C(t)=\nu \langle p^2(t)\rangle$. The physics behind the quantized response of $C(t)$ is the quantized acceleration of a soliton in the momentum space. Specifically, the $\cal{PT}$-symmetric driving potential induces the formation of a soliton, which is localized in $\theta_c=\pi/2$ and moves unidirectionally in the momentum space with the quantized rate, i.e., $\langle p \rangle=Dt$, where $D=2m\pi$ with arbitrary $K\in (2m\pi-\pi,2m\pi+\pi)$.

The exotic quantization of OTOCs broadens our understanding on the chaotic dynamics in the presence of non-Hermitian effects, and is of great significance in the updated experiments in ultracold atoms and photonics.
So far, several experiments in the superconducting qubits~\cite{Dressel18,Alonso19}, traped irons~\cite{Joshi20}, and NMR~\cite{Li17,Nie20} have realized
the measurement of OTOCs. However, the research finding of $C(t)=\nu\langle p^2(t)\rangle$ blazes a new trail in measuring OTOCs by using the
mean square of the momentum, which makes the observation of OTOCs much easier to realize in experiments.
We then propose an experimental scheme to measure the quantization of OTOCs.
It is known that, under the paraxial approximation, the propagation of light is governed by Schr\"odinger-like equation, with the longitude dimension of light propagation mimicking the time variable~\cite{Sharabi,Prange89}.
Based on this, the kicked rotor model can be realized by using the optical setting, where the light propagates in periodical arrays of phase gratings in space~\cite{Rosen00} and thus mimics the  quantum kicked rotor subjected periodical kicks in time domain. The $\cal{PT}$-symmetric potential is introduced by the effects of sinusoidal and quarter-wave-shifted gratings~\cite{Longhi17}. The mean square of momentum is measured in the frequency domain of the light.  Therefore, our finding of the quantized dynamics of OTOCs is within the reach of updated experiments.

{\color{blue}\textit{Acknowledgements.--}}
I am grateful to Jie Liu, Zhi Li, and Longwen Zhou for valuable suggestions
and discussions. This work was supported by the Natural Science Foundation of China under Grant No. 12065009.

\pagebreak
\clearpage
\begin{center}
\textbf{\large Supplemental Material:\\ Quantization of Out-of-time-ordered correlators in non-Hermitian chaotic systems}
\end{center}
\setcounter{equation}{0} \setcounter{figure}{0} \setcounter{table}{0}
\setcounter{page}{1} \makeatletter \global\long\def\theequation{S\arabic{equation}}
 \global\long\def\thefigure{S\arabic{figure}}
 \global\long\def\bibnumfmt#1{[S#1]}
 \global\long\def\citenumfont#1{S#1}

\section*{S1. Details about numerical methods to calculate $C(t_n)$}
As shown in the main text, the OTOCs can be decomposed as
\begin{align}\label{Decomposition-SM}
C(t_n) & =  C_1(t_n) + C_2(t_n) - 2{\rm{Re}}[C_3(t_n)]\;,
\end{align}
where the first term
\begin{align}\label{FPart-SM}
C_{1}(t_n)& \mathrel{\mathop:}= \langle p^{\dagger}(t_n)\theta^2p(t_n) \rangle=\langle \psi_R(t_0) | \theta^2 |\psi_R(t_0)\rangle\;,
\end{align}
the second one
\begin{align}\label{SPart-SM}
C_{2}(t_n) &\mathrel{\mathop:}= \langle \theta p^{\dagger}(t_n)p(t)\theta \rangle =\langle \varphi_R(t_0)|\varphi_R(t_0)\rangle
\end{align}
and the third one
\begin{align}\label{TPart-SM}
C_{3}(t_n) &\mathrel{\mathop:}= \langle p^{\dagger}(t_n)\theta p(t_n)\theta \rangle=\langle \psi_R(t_0)|\theta |\varphi_R(t_0)\rangle\;,
\end{align}
with $|\psi_R(t_0)\rangle = U^{\dagger}(t_0,t_n)pU(t_0,t_n)|\psi(t_0)\rangle$ and $|\varphi_R(t_0)\rangle =U^{\dagger}(t_0,t_n)pU(t_0,t_n)\theta|\psi(t_0)\rangle$ being the states at the end of time reversal.

Numerical method for calculating $C_{1}(t_n)$ of a specific $t_n$ is as following [see Table.~\eqref{SchemDigm}],
\begin{itemize}
\item[i).] Prepare a state $|\psi (t_0)\rangle$ at the initial time $t_0=0$.
\item[ii).] Conduct forward evolution from $t_0$ to $t_n$. The state $|\psi(t_n)\rangle = U(t_0,t_n)|\psi (t_0)\rangle$ can be obtained.\\
    Note that, for non-Hermitian systems, the norm $\mathcal{N}(t_j) = \langle \psi(t_j)|\psi(t_j)\rangle$ will exponentially increases with time. To drop the contribution of the norm to OTOCs, we take normalization for the time-evolved state, namely, $|\psi(t_j)\rangle =|\psi(t_j)\rangle\sqrt{\mathcal{N}(t_0)/\mathcal{N}(t_j)}$ with $t_0<t_j\leq t_n$ and $\mathcal{N}(t_j)= \langle \psi(t_j)|\psi(t_j)\rangle$ denotes the norm before normalization. As a consequence, the norm equals to that of the initial state, i.e., $\mathcal{N}(t_j)=\mathcal{N}(t_0) = \langle\psi (t_0)|\psi (t_0)\rangle=1$ [see Table.~\eqref{SchemDigm}].
\item[iii).] Exert the operator $p$ on $|\psi(t_n)\rangle$ at the time $t_n$. One can obtain a new state $|\tilde{\psi} (t_n)\rangle = p |\psi(t_n)\rangle$. \\
    Note that, the norm of $|\tilde{\psi} (t_n)$  is just the mean square of momentum at the time $t=t_n$, i.e., $\tilde{\mathcal{N}}_{\psi} (t_n) =\langle \tilde{\psi}(t_n)|\tilde{\psi}(t_n)\rangle =\langle \psi(t_n)|p^2|\psi(t_n)\rangle$.
\item[iv).] Conduct backward evolution from
$t_n$ to $t_0$. The state $|\psi_R(t_0)\rangle = U^{\dagger}(t_0,t_n)|\tilde{\psi} (t_n)\rangle = U^{\dagger}(t_0,t_n)p |\psi(t_n)\rangle$ can be obtained.\\ Same as in step ii), we take normalization for the time-reversed state $|\psi_R(t_j)\rangle = |\psi_R(t_j)\rangle \sqrt{{\tilde{\mathcal{N}}_{\psi}(t_n)}/{\mathcal{N}_{\psi_R}(t_j)}}$ with $t_0\leq t_j<t_n$ and $\mathcal{N}_{\psi_R}(t_j)= \langle \psi_R(t_j)|\psi_R(t_j)\rangle$ denotes the norm before normalization. So the value of  $\tilde{\mathcal{N}}_{\psi}(t_n)$ stays unchanged, i.e., $\tilde{\mathcal{N}}_{\psi}(t_n)=\mathcal{N}_{\psi_R}(t_0)$ [see Table.~\eqref{SchemDigm}].
\item[v).] At the end of time reversal $t=t_0$, calculate the average of the operator $\theta^2$, and $C_{1}(t_n)=\langle \psi_R(t_0)| \theta^2 |\psi_R(t_0)\rangle$ [see Eq.~\eqref{FPart-SM}].
\end{itemize}

For the numerical procedure to calculate  $C_2(t_n)$, there are also five steps:
\begin{itemize}
\item[i).] Prepare a state $|\psi (t_0)\rangle$ at the initial time $t_0=0$. Exert the operator $\theta$ on this state. A new state $|\varphi(t_0)\rangle=\theta |\psi (t_0)\rangle$ can be obtained. The norm is $\mathcal{N}_{\varphi_0} = \langle\varphi(t_0)|\varphi(t_0)\rangle =\langle\psi (t_0)|\theta^2|\psi (t_0)\rangle$, which is just the mean square of angle coordinate of the initial state $|\psi(t_0)\rangle$.
    \item[ii)]\hspace*{-1.5mm}-iv). Same as in calculating $C_1(t_n)$.
\item[v).] Calculate the norm of the state $|\varphi_R(t_0)\rangle$ at the end of time reversal $t=t_0$, i.e., $\mathcal{N}_{\varphi_R}(t_0)=\langle \varphi_R(t_0)|\varphi_R(t_0)\rangle$ which is just the $C_{2}(t_n)$ [see Eq.~\eqref{SPart-SM}].
\end{itemize}
At the end of time reversal, we use the states $|\psi_R(t_0)\rangle$ and $|\varphi_R(t_0)\rangle$
to calculate the four-point correlator $C_3(t)= \langle \psi_R(t_0)|\theta |\varphi_R(t_0)\rangle$ [see Eq.~\eqref{TPart-SM}].

\begin{table*}[htbp]
\begin{center}
\begin{tabular}{|p{3cm}<{\centering}|c|c|c|c|c|c|c|}
\hline
 & \multicolumn{3}{|c|}{Forward: $t_n$ steps } &$p$ action &\multicolumn{3}{|c|}{Backward: $t_n$ steps }\\
\cline{2-8}
\rule{0pt}{12pt} &
\multicolumn{3}{|l|}{ $|\psi(t_0)\rangle $ $\rightarrow$ $|\psi(t_1)\rangle$ $\rightarrow$ $\cdots$ $|\psi(t_n)\rangle$} &$|\tilde{\psi}(t_n)\rangle= p |\psi(t_n)\rangle$ &
\multicolumn{3}{|l|}{$|\tilde{\psi}(t_n)\rangle$ $\cdots$ $\rightarrow$ $|\psi_R(t_1)\rangle$ $\rightarrow$ $|\psi_R(t_0)\rangle$}\\
\hline
\rule{0pt}{12pt}$Q(t)=\langle\psi(t)|\theta^2|\psi(t)\rangle$ & \multicolumn{3}{|l|}{\hspace*{0mm}$Q(t_0) $\hspace*{3mm}$\rightarrow$\hspace*{2mm}$Q(t_1)$ \hspace*{0mm} $\rightarrow$ \hspace*{0mm}$\cdots$\hspace*{0mm}$Q(t_n)$} &$\tilde{Q}(t_n)=\langle \tilde{\psi}(t_n)|\theta^2|\tilde{\psi}(t_n)\rangle$ &\multicolumn{3}{|l|}{\hspace*{1mm}$\tilde{Q}(t_n)$\hspace*{2mm}$\cdots$ \hspace*{-1mm} $\rightarrow$\hspace*{0.8mm} $Q_R(t_1)$\hspace*{0mm} $\rightarrow$\hspace*{1mm}$Q_R(t_0)$}\\
\hline
\rule{0pt}{12pt}$E(t)=\langle\psi(t)|p^2|\psi(t)\rangle$ & \multicolumn{3}{|l|}{\hspace*{0mm}$E(t_0) $\hspace*{3mm}$\rightarrow$\hspace*{2mm}$E(t_1)$ \hspace*{0mm} $\rightarrow$ \hspace*{0mm}$\cdots$\hspace*{0mm}$E(t_n)$} &$\tilde{E}(t_n)=\langle \tilde{\psi}(t_n)|p^2|\tilde{\psi}(t_n)\rangle$ &\multicolumn{3}{|l|}{\hspace*{1mm}$\tilde{E}(t_n)$\hspace*{2mm}$\cdots$ \hspace*{-1mm} $\rightarrow$\hspace*{0.8mm} $E_R(t_1)$\hspace*{0mm} $\rightarrow$\hspace*{1mm}$E_R(t_0)$}\\
\hline
\rule{0pt}{12pt}$\mathcal{N}(t)=\langle\psi(t)|\psi(t)\rangle$ & \multicolumn{3}{|l|}{\hspace*{0mm}$\mathcal{N}(t_0) $\hspace*{2mm}$=$\hspace*{2mm}$\mathcal{N}(t_1)$ \hspace*{0mm} $=$ \hspace*{1mm}$\cdots$\hspace*{1mm}$\mathcal{N}(t_n)=1$} &$\tilde{\mathcal{N}}(t_n)=\langle \psi(t_n)|p^2|\psi(t_n)\rangle$ &\multicolumn{3}{|l|}{\hspace*{1mm}$\tilde{\mathcal{N}}(t_n)$\hspace*{1mm}$\cdots$ \hspace*{0.1mm}  $=$\hspace*{1mm}${\cal{N}}_R(t_1)$\hspace*{1mm} $=$\hspace*{1mm}${\cal{N}}_R(t_0)$}\\
\hline
\end{tabular}
\caption{Schematic diagram of time evolution to calculate the term $C_1(t_n)=Q_R(t_0)=\langle\psi_R(t_0)|\theta^2|\psi_R(t_0)\rangle$. }\label{SchemDigm}
\end{center}
\end{table*}

\end{document}